\documentclass[a4paper,reqno,11pt]{amsart}
\usepackage{amssymb,latexsym}

\begin{document}

\title[Doubled Field Approach to Yang - Mills Requires
Non-Locality]
{Doubled Field Approach to Yang - Mills Requires
Non-Locality}

\author{Alexei~J. Nurmagambetov}

\address{A.I. Akhiezer Institute for Theoretical Physics\\
NSC ``Kharkov Institute of Physics and Technology"\\
Kharkov, 61108, Ukraine}

\email{ajn@kipt.kharkov.ua}
\date{\today}

\begin{abstract}
Doubling a Yang-Mills field we apply the pattern which has been
found to construct a ``duality-symmetric" gravity with matter to
the ``duality-symmetric" Yang - Mills theory in five space-time
dimensions. Constructing the action we conclude that dualizing a
non-abelian theory requires non-locality. We analyze the
symmetries of the theory and equations of motion. Extension to the
supersymmetric theory is also demonstrated.

\end{abstract}

\maketitle

\vspace{1.3cm}

Some of the fields of Superstring/M-theory spectrum are of special
class which are called chiral p-forms, or chiral bosons. These
fields play an important r\^ole in establishing various dualities
between different sectors of M-theory, but dealing with them
beyond the mass-shell, i.e. at the level of the effective
Lagrangians, is not a simple task. There are several methods (see
\cite{sorokin} for a review and for the comprehensive list of
Refs.) which were proposed to describe theories with self-dual or
duality-symmetric fields. All of them could be split into the
following major sets. The first one \cite{z}-\cite{ht} contains
the approaches that are duality invariant but are not manifestly
Lorentz invariant. Introducing auxiliary fields is not required,
but coupling to other fields, especially to gravity, may cause
problems with establishing the consistency of such a coupling. The
second set \cite{s}-\cite{pst} is dealt with auxiliary fields
whose inclusion restores the Lorentz covariance. The number of
these auxiliary fields may vary from one to infinity.

Among the approaches with auxiliary fields the formalism proposed
by Pasti, Sorokin and Tonin \cite{pst} takes a special place. It
is manifestly Lorentz covariant and is minimal in a sense of
having the only auxiliary field entering the action in a
non-polynomial way. Successful applying the PST approach to the
construction of different field theories of chiral p-forms,
super-p-branes with worldvolume chiral fields, and of different
sub-sectors of supergravities has demonstrated the advantages of
this approach and its compatibility with supersymmetry (cf.
\cite{sorokin} and Refs. therein). However, a gap for applying the
PST formalism is a Yang-Mills theory.

It is worth noting that the problem of dualising a non-abelian
gauge theory has been intensively studied in literature. Getting
rid of self-interactions it is straightforward to apply the
machinery of dualization to the case. But the attempts to go
beyond the free theory have faced the troubles. The latter can be
summarized in the ``no-go" theorems \cite{dt}, \cite{gy} which
forbid a trivial generalization of the well-known
electric-magnetic duality of Maxwell theory. The point is that the
Poincar\'{e} lemma does not directly generalize when dealing with
a YM covariant derivative that in its turn prevents the
straightforward applying the PST formalism to formulate a
duality-symmetric YM theory \cite{bcrep}, \cite{bc}.

Searching for a dualization of a non-abelian theory becomes now
important from the point of view of pushing forward the doubled
field approach of \cite{cjlp0}, \cite{cjlp} to the non-maximal
supergravities. On the supergravity side much has been done in the
dualization program for maximal supergravities in diverse
dimensions \cite{cjlp0}, \cite{cjlp} that can be obtained from
D=11 or D=10 IIA/B supergravities by toroidal dimensional
reduction. There doubling the fields of the gauge sector of
supergravities it has been demonstrated that the original
equations of motion of a theory admit the representation in terms
of the Bianchi identities for the dual fields, and moreover, the
dynamical content of a theory is encoded into the so-called
twisted self-duality condition relating the original and the dual
field strengths. The way to go beyond the mass-shell by lifting
the approach of \cite{cjlp} onto the level of the proper action
was proposed for D=11 and D=10 type IIA supergravities in
\cite{bns}, \cite{nur}. However, the doubled field approach can
not be directly applied for the non-maximal supergravities in D=10
and for the low-dimensional gauged supergravities where the
non-abelian fields become a part of the supergravity multiplet.

Therefore, to realize the aforementioned program we are forced to
figure out a way to deal with non-abelian fields. One of the ways
is to find a generalization of the Hodge star notion to a
non-abelian case. Such a generalization has been proposed in
\cite{cft} and requires essentially non-local consideration since
it is based on a loop space formulation of a gauge theory. We are
aimed to reach the same conclusion on the non-local character of a
dualization of a non-abelian theory being on the ground of
standard approach to the Yang-Mills theory. Since the YM theory
possesses the same as the gravity theory feature of having a
self-interaction we will use this fact to establish the properties
of the doubled field approach to the (S)YM theory which will help
us in pushing forward the same approach to the (super)gravity
case. And since we will mostly interested in a dualization of D=11
(super)gravity, in a spirit of recent studying a hidden symmetry
group of M-theory \cite{westjhep00}-\cite{chaud}, we will exploit
the tight relation between five-dimensional simple supergravity
and D=11 supergravity \cite{cremmer} to study the YM theory in
D=5.

As a first step in recovering the doubled field action for the YM
theory we have to find, without an appeal to a method of
constructing such an action, a convenient representation of the YM
equation of motion in a way that allows us to present the latter
as the Bianchi identity for a dual field. It turns out to be
convenient to write down the YM equation of motion in a form which
is very similar to the dynamics of Maxwell theory with an
electric-type source. Such a representation suggests a way of
extracting the dual field after that we can apply the machinery of
the PST approach to construct the action from which the duality
relations between the YM field and its dual partner will follow as
equations of motion.

To do so let us get started with the following action for a
Yang-Mills theory in D=5
\begin{equation}\label{YM}
S_{YM}=-\frac{1}{2}{\mbox Tr}\int_{{\mathcal M}^5}\, {F}^{(2)}
\ast {F}^{(2)},
\end{equation}
where ${F}^{(2)}\equiv D{A}^{(1)}=d{A}^{(1)}-{1\over 2}ig [{
A}^{(1)},{A}^{(1)}]$ is the field strength in the adjoint
representation of a semi-simple non-abelian group, $g$ is a
coupling constant, and the wedge product between forms has to be
assumed. The action $S_{YM}$ and the gauge fields equation of
motion
\begin{equation}\label{YMem}
D\ast { F}^{(2)}=0
\end{equation}
are invariant under the local non-abelian gauge transformations
\begin{equation}\label{gtr}
\delta { A}^{(1)}={1\over g}D{ \alpha}^{(0)}\equiv {1\over g}d{
\alpha}^{(0)}+i[{ A}^{(1)},{ \alpha}^{(0)}].
\end{equation}

To apply the Poincar\'{e} lemma let us present the equation of
motion \eqref{YMem} in a slightly different form extracting the
part with usual, non-covariant, derivative, and separating the
free part from that of describing the self-interaction of a YM
field. Taking into account the definition of the YM field strength
we get
\begin{equation}\label{eom}
d(\ast d{ A}^{(1)})=\ast { J}^{(1)}
\end{equation}
with
\begin{equation}\label{J}
\ast { J}^{(1)}=ig[{ A}^{(1)}, \ast { F}^{(2)}]+d\ast\left({1\over
2}ig[{ A}^{(1)},{ A}^{(1)}]\right).
\end{equation}
Since $d^2=0$, in the trivial topology setting one could notice
\begin{equation}\label{G}
\ast {J}^{(1)}=d \ast { G}^{(2)},
\end{equation}
where ${ G}^{(2)}$ is a function of the YM potentials ${ A}^{(1)}$
and their derivatives and as we will see in what follows is the
source of non-locality since formally we can resolve \eqref{G}
through the non-local expression
\begin{equation}\label{GJ}
\ast{ G}^{(2)}=d^{-1}\ast { J}^{(1)}.
\end{equation}
Here we have introduced the inverse to $d$ operator whose action
can be understood as follows. Let us introduce
a ``Green" function to the equation
\begin{equation}\label{h}
dh(x)=\delta^5 (x),
\end{equation}
with the Dirac delta-function on the r.h.s. Then
we define the action of $d^{-1}$
on an arbitrary form at a space-time point $x$ as
\begin{equation}\label{d-1}
d^{-1}(x)\omega^{(p)}(x)=(-)^p \int d^5 y\, h(x-y)~
\omega^{(p)}(y).
\end{equation}
To make a sense the latter expression should only deal with the
``Green" functions that act on a causality-related space-time
region.

The dual to the one-form ${ J}^{(1)}$ is a conserved ``current"
which is the YM analog of the gravity Landau-Lifshitz
pseudo-tensor, and eq. \eqref{eom} is an analog of the equation of
motion of Maxwell field with an electric-type current. Note that
the ``current" entering the r.h.s. of \eqref{eom} is not gauge
invariant, but since the l.h.s. of the same equation is not gauge
invariant too, the latter compensates the former leaving eq.
\eqref{eom} to be invariant under the local gauge transformations
\eqref{gtr}.

Having the representation \eqref{eom} we can double the YM field
with its ``dual" partner and write down this equation of motion as
the Bianchi identity for the YM ``dual"
\begin{equation}\label{defB}
d{ B}^{(2)}=\ast (d{ A}^{(1)} - { G}^{(2)}),
\end{equation}
or equivalently
\begin{equation}\label{calF3}
{\mathcal F}^{(3)}=0,\qquad {\mathcal F}^{(3)}=d{ B}^{(2)}-\ast
(d{ A}^{(1)} - { G}^{(2)}).
\end{equation}
Indeed, applying the operator $d$ to \eqref{defB} or \eqref{calF3}
leads to the YM equation of motion \eqref{YMem}
\begin{equation}\label{dcalF3}
d{\mathcal F}^{(3)}=D\ast { F}^{(2)}=0.
\end{equation}
Therefore, an equivalent way of a description of a YM theory is to
find the action from which eq. \eqref{calF3} will follow as an
equation of motion.

However, the action we shall construct should be gauge invariant
as well as equations of motion which will follow from that action.
Therefore, we get to inspect the gauge invariance more closer. To
this end let us recall that the action of the local gauge
transformations \eqref{gtr} on the YM field strength results in
the rotation of the latter in a group space, i.e. $\delta {
F}^{(2)}=-i[{ \alpha}^{(0)},{ F}^{(2)}]$. To find the similar
transformation law for the ${\mathcal F}^{(3)}$ it is convenient
to present \eqref{eom} as
\begin{align}\label{eom1}
d\ast { F}^{(2)}=&\ast \tilde{ J}^{(1)},\notag\\
&\ast\tilde{ J}^{(1)}=ig[{ A}^{(1)}, \ast { F}^{(2)}]=d\ast\tilde{
G}^{(2)}.
\end{align}

Then, using the $F^{(2)}$ gauge transformation law one can derive
from \eqref{eom1} the transformation of $\ast\tilde{ G}^{(2)}$
\begin{equation}\label{Ggtr}
\delta\ast\tilde{ G}^{(2)}=-i[\alpha^{(0)},{\mathcal F}^{(3)}+\ast
F^{(2)}]-d^{-1}\left( i[d\alpha^{(0)},{\mathcal F}^{(3)}] \right),
\end{equation}
which is the non-local gauge transformation in view of the
non-local character of this quantity.

To require $\delta {\mathcal F}^{(3)}=-i[{ \alpha}^{(0)},{\mathcal
F}^{(3)}]$ one has to assign the following non-local non-abelian
gauge transformation to the ${ B}^{(2)}$ field
\begin{equation}\label{Bgtr}
\delta{ B}^{(2)}=d{ \alpha}^{(1)}+d^{-2}\left(
i[d\alpha^{(0)},{\mathcal F}^{(3)}]\right).
\end{equation}
What concerns to the gauge invariance of the equations of motion,
we should emphasize that the standard YM equation of motion
\eqref{YMem} is only on-shell invariant under the action of
\eqref{gtr} since $\delta (D\ast { F}^{(2)})=-i[{ \alpha}^{(0)},
D\ast { F}^{(2)}]$. The same concerns to the $d{\mathcal
F}^{(3)}=0$ since this expression is gauge invariant only on the
shell of the duality relation ${\mathcal F}^{(3)}=0$.

Therefore, our attempt to stay on the ground of applying usual
Poincar\'{e} lemma to the non-abelian case has faced the necessity
of dealing with non-locality due to the ``source"-like terms which
appear in the non-abelian gauge field equation of motion after
fitting the later for the application of the Poincar\'{e} lemma.
It is easy to see that this non-localities disappear in the zero
gauge coupling constant limit $g\rightarrow 0$ when ${
G}^{(2)}\rightarrow 0$, ${\mathcal F}^{(3)}\rightarrow d{ B}-\ast
d{ A}$, $\delta{ A}=d\tilde{ \alpha}^{(0)}$, $\delta{ B}^{(2)}=d{
\alpha}^{(1)}$, and therefore we are effectively dealing with
${\mathcal N}$ copies of the abelian duality-symmetric fields
where ${\mathcal N}$ is the dimension of the non-abelian group.
However, it does not contradict with the ``no-go" theorem of
\cite{bc} since the extention of a system of ${\mathcal N}$ copies
of free duality-symmetric abelian fields to a non-abelian system
comes through introducing the non-local quantities. The other
feature of the construction that has to be noticed consists in a
non-equivalence of the original YM field and its ``dual" partner
since the non-abelian extension of the latter is formed by the
part containing the self-interaction of the former. Indeed, the
equation of motion for the dual field is
\begin{equation}\label{Beom}
d(\ast dB^{(2)})=-dG^{(2)},
\end{equation}
that follows from the duality relation $\ast{\mathcal F}^{(3)}=0$.
However, one can essentially simplify this equation with taking
into account the Hodge identity
\begin{equation}\label{Hid}
d\triangle^{-1} {\mathbf\delta}+{\mathbf\delta}\triangle^{-1} d=1,
\end{equation}
where ${\mathbf\delta}$ is the co-derivative and $\triangle^{-1}$
is the inverse to the Laplacian
$\triangle=d{\mathbf\delta}+{\mathbf\delta}d$ operator. Using the
Hodge identity one can present $\ast{\mathcal F}^{(3)}=0$ as
\begin{equation}\label{calF2m}
\ast dB^{(2)}=F^{(2)}-\ast \triangle^{-1}{\mathbf\delta}\ast
\tilde J^{(1)},
\end{equation}
and since the last term on the r.h.s. of the latter equation is a
closed form, the equation of motion of $B^{(2)}$ is
\begin{equation}\label{Beom1}
d(\ast dB^{(2)})=-\frac{1}{2}ig d\left([A^{(1)},A^{(1)}]\right).
\end{equation}
Therefore, there is not a symmetry similar to the symmetry under
duality rotations in Maxwell theory that is closely related to the
``no-go" theorem of \cite{dt}.

Let us now turn to the construction of the action from which the
duality relation ${\mathcal F}^{(3)}$ will follow as an equation
of motion. Taking into account an analogy with the gravity case
considered in \cite{ajngr} it is quite naturally to guess the
following term
\begin{equation}\label{PST}
S_{PST}={1\over 2}{\mbox Tr}\int_{{\mathcal M}^5}\, v~{\mathcal
F}^{(3)}~i_v {\mathcal F}^{(2)},
\end{equation}
as the main candidate that has to be added to the action
\eqref{YM}. Here the one-form $v$ is constructed out of the PST
scalar field $a(x)$ ensuring the covariance of the model
\begin{equation}\label{v}
v={d a(x)\over {\sqrt{-(\partial a)^2}}},
\end{equation}
${\mathcal F}^{(3)}$ has appeared in \eqref{calF3}, and
\begin{equation}\label{calF2}
{\mathcal F}^{(2)}=d{ A}^{(1)}-\ast(d{ B}^{(2)}+\ast {
G}^{(2)}),\quad {\mathcal F}^{(3)}=-\ast {\mathcal F}^{(2)}.
\end{equation}
It is clear from the previous discussion that the generalized
field strengths \eqref{calF3} and \eqref{calF2} are the covariant
under the gauge transformations objects, although the quantities
entering them are not covariant and the action \eqref{PST} is
invariant under the gauge transformations \eqref{gtr},
\eqref{Ggtr}, \eqref{Bgtr} which leave the PST scalar intact.

To prove the relevance of the proposed term, let us consider a
general variation of \eqref{PST}. The standard manipulations (see
\cite{bns} for details) result in
\begin{align}\label{vPST}
\delta {\mathcal L}_{PST}&={\mbox Tr}\left( \delta {
B}^{(2)}+{\delta a \over {\sqrt{-(\partial a)^2}}} i_v {\mathcal
F}^{(3)} \right)~d(v~i_v{\mathcal F}^{(2)})\notag\\& -{\mbox
Tr}\left( \delta { A}^{(1)}+{\delta a \over {\sqrt{-(\partial
a)^2}}} i_v {\mathcal F}^{(2)} \right)~ d(v~i_v{\mathcal
F}^{(3)})\notag\\ &-{\mbox Tr}~ \delta (\ast {
G}^{(2)})~v~i_v{\mathcal F}^{(2)}-{\mbox Tr}~\delta { A}^{(1)}~
d{\mathcal F}^{(3)},
\end{align}
where we have omitted the total derivative term.

The last term of \eqref{vPST} is precisely the term whose
contribution is cancelled against the variation of $S_{YM}$.
Therefore, the complete action $S=S_{YM}+S_{PST}$ is invariant
under the non-abelian gauge transformations and the following two
sets of special symmetry \cite{pst}
\begin{align}\label{pst1}
\delta a(x)=0,\qquad &\delta { A}^{(1)}=da~{ \varphi}^{(0)},\notag\\
&(d\delta{ B}^{(2)}+\delta\ast{
G}^{(2)})=da~d{ \varphi}^{(1)} \Longrightarrow \notag\\
&\delta { B}^{(2)}=da~{ \varphi}^{(1)}-d^{-1}\delta(\ast{
G}^{(2)}),
\end{align}
\begin{align}\label{pst2}
\delta a(x)=\Phi (x),\quad &\delta { A}^{(1)}=-{\delta a \over
{\sqrt{-(\partial a)^2}}} i_v{\mathcal F}^{(2)},\notag\\ &\delta {
B}^{(2)}=-{\delta a \over {\sqrt{-(\partial a)^2}}} i_v{\mathcal
F}^{(3)}-d^{-1}\delta(\ast{ G}^{(2)}).
\end{align}

Let us now discuss how these special symmetries do the job. The
equations of motion of ${ B}^{(2)}$ and ${ A}^{(1)}$ that follow
from the action $S=S_{YM}+S_{PST}$ are
\begin{equation}\label{B2eom}
d(v~i_v{\mathcal F}^{(2)})=0,
\end{equation}
\begin{equation}\label{A1eom}
d(v~i_v{\mathcal F}^{(3)})+{{\mbox Tr}(v~i_v{\mathcal
F}^{(2)}~\delta\ast { G}^{(2)}) \over \delta { A}^{(1)}}=0,
\end{equation}
where we have used that $\ast { G}^{(2)}$ is a function of the YM
potentials ${ A}^{(1)}$.

The general solution to the equation of motion \eqref{B2eom} is
\cite{pst}
\begin{equation}\label{solB2}
v~i_v{\mathcal F}^{(2)}=da~d{ \xi}^{(0)}.
\end{equation}
Using the symmetry \eqref{pst1} with ${ \varphi}^{(0)}={
\xi}^{(0)}$ one can obtain from \eqref{solB2}
\begin{equation}\label{1solB2}
i_v{\mathcal F}^{(2)}=0 \quad \rightsquigarrow \quad {\mathcal
F}^{(2)}=0.
\end{equation}
Taking the latter into account and using the same trick one can
obtain from \eqref{A1eom}
\begin{equation}\label{solA1}
i_v{\mathcal F}^{(3)}=0 \quad \rightsquigarrow \quad {\mathcal
F}^{(3)}=0.
\end{equation}
It becomes clear that equation of motion of the PST scalar $a(x)$
\begin{equation}\label{aeom}
{\mbox Tr}\left( i_v{\mathcal F}^{(3)}~d(v~i_v{\mathcal F}^{(2)})-
i_v{\mathcal F}^{(2)}~d(v~i_v{\mathcal F}^{(3)})\right)=0
\end{equation}
does not contain a new dynamical information and is satisfied
identically as a consequence of the equations of motion
\eqref{B2eom}, \eqref{A1eom}. Indeed, eq. \eqref{aeom} is the
Noether identity which is a reflection of a local symmetry which
is nothing but the symmetry under \eqref{pst2}.

Therefore, we have proved that the action $S=S_{YM}+S_{PST}$ is
the one we are looking for. The action possesses the special
symmetries \eqref{pst1}, \eqref{pst2} which have to be used to
derive from equations of motion the duality relations between the
YM field and its dual partner and to establish the auxiliary
nature of the PST scalar field. Owing to the symmetry \eqref{pst2}
the PST scalar field does not spoil the original content of a
theory and is the pure auxiliary field.

To extend this construction to the supersymmetric case recall that
the supersymmetric counterpart of the D=5 YM theory is described
by
\begin{equation}\label{SYM}
S_{SYM}=-{1\over 2}~{\mbox Tr} \int_{{\mathcal M}^5}\, \left(
F^{(2)}\ast F^{(2)}+i\bar{\lambda}\Gamma^a D{ \lambda} \Sigma_a
+\dots\right),
\end{equation}
where the four-form $\Sigma_a$ is defined by
\begin{equation}\label{Sig}
\Sigma_a={1\over 4!}\epsilon_{abcde} E^{b} E^{c} E^{d} E^{e}
\end{equation}
with the vielbeins $E^a$, and $D{ \lambda}$ is the covariant
derivative of the gaugino field. Note that we have kept only the
terms essential for the consideration in what follows neglecting
the scalars and auxiliary fields which cast the off-shell N=2
Yang-Mills supermultiplet in D=5.

The action \eqref{SYM} is in particular invariant under the
following global supersymmetry transformations
\begin{equation}\label{susyYM}
\delta_{\epsilon}{ A}^{(1)}=-{i\over 2}\bar{\epsilon}\Gamma^{(1)}
{ \lambda},\qquad \delta_{\epsilon}{ \lambda}={1\over 2}\ast (\ast
{ F}^{(2)}~\Gamma^{(2)})\epsilon,
\end{equation}
where we have used the following notation for gamma-matrices
\begin{equation}\label{Gn}
\Gamma^{(n)}={1\over n!} E^{a_n}~\dots~E^{a_1} \Gamma_{a_1\dots
a_n}.
\end{equation}

To find the appropriate supersymmetry transformations for the
doubled field version of the super-Yang-Mills theory it is
convenient to present the PST part of the action as
\begin{equation}\label{PSTSYM}
S_{PST}=-{1\over 2}~{\mbox Tr} \int_{{\mathcal M}^5}\,
i_v{\mathcal F}^{(2)}\ast i_v{\mathcal F}^{(2)}
\end{equation}
with
\begin{equation}\label{calF2m1}
{\mathcal F}^{(2)}={ F} ^{(2)}-\ast(d{
B}^{(2)}+\ast\mathcal{G}^{(2)}),
\end{equation}
where $\mathcal{G}^{(2)}$ is the extention of $\tilde{ G}^{(2)}$
with a non-local term coming from the fermionic current.

Then it is easy to verify that the action $S=S_{SYM}+S_{PST}$ is
invariant under the following global supersymmetry transformations
\begin{align}\label{susyaA}
\delta_{\epsilon} a=0,\qquad &\delta_{\epsilon}{ A}^{(1)}=-{i\over
2}\bar{\epsilon}\Gamma^{(1)} { \lambda},\notag\\
&\delta_{\epsilon}{ \lambda}={1\over 2}\ast \left(\ast [{
F}^{(2)}+v~i_v{\mathcal F}^{(2)}]~\Gamma^{(2)}\right)\epsilon.
\end{align}
The supersymmetry transformation of the dual to the YM field can
be recovered from the requirement
\begin{equation}\label{susyB}
\delta_{\epsilon}\left( \ast d{ B}^{(2)}+\mathcal{G}^{(2)}
\right)=0.
\end{equation}
Hence, the proposed extension of the PST technique to a
non-abelian case is compatible with supersymmetry but requires the
non-local terms in the supersymmetry transformation of the dual
field.

To summarize, we have presented the Yang-Mills equation of motion
in the form which is very likely to that of Maxwell theory with an
electric-type current. The YM ``current" form so obtained encodes
the self-interaction between the Yang-Mills fields, does not
possesses the local gauge invariance and is a closed form. The
latter allows one to present the ``current" form as a curl of a
``current potential" and therefore to rewrite the second order YM
equation of motion as the first order Bianchi identity for the
dual field. Since the form of the ``current potential" is defined
by the non-local expression we have demonstrated  that such a
dualization of the Yang-Mills requires non-locality. But the
latter does not spoil the general scheme of constructing
duality-symmetric theories \`{a} la PST, though the trace of the
non-locality can be observed in the gauge transformations of the
doubled field YM action.

The same story happens in the gravity case \cite{ajngr}. Indeed,
after resolving the torsion free constraint in the first order
formulation of the Einstein-Hilbert action, one can present the
gravity equation of motion in a similar to the eq. \eqref{eom}
form. Therefore in framework of the standard approach dualizing
the gravity requires introducing non-localities too. However, we
have mentioned above that there is an alternative way of
non-abelian generalization of electric-magnetic duality based on
the loop space formulation of a gauge theory in D=4 space-time
dimensions \cite{cft}. An analog of such a formulation of
gravitational theory is nothing but the Ashtekar-Sen approach (see
e.g. \cite{smolin}, \cite{thie} for reviews). It would be
interesting to figure out how the loop space approach could be
reformulated to describe a duality-symmetric theory where the
original and the dual potentials will appear on equal footing.

We are very grateful to Igor Bandos and Dmitri Sorokin for
valuable comments, suggestions and encouragement. This work is
supported in part by the Grant \# F7/336-2001 of the Ukrainian
SFFR and by the INTAS Research Project \#2000-254.

\end{document}